\documentclass[12pt]{article}
\usepackage{graphicx,amsmath,amssymb}
\usepackage{indentfirst}
\usepackage[usenames]{color}
\usepackage[colorlinks=true, urlcolor=navyblue, linkcolor=navyblue, citecolor=navyblue]{hyperref}
\usepackage{epstopdf}
\usepackage{enumerate}
\usepackage{appendix}
\usepackage{lineno}
\usepackage{setspace}

\definecolor{navyblue}{rgb}{0,0.08,0.45}

\begin{document}


\begin{flushright}
{\small JLAB-PHY-16-2198\\
SLAC--PUB--16449 \\ \vspace{2pt}}
\end{flushright}

\vspace{0pt}

\begin{center}
{\huge  On the Interface between Perturbative}

\vspace{10pt}

{\huge   and Nonperturbative QCD}

\end{center}

\vspace{10pt}

\centerline{Alexandre Deur}

\vspace{3pt}

\centerline{\it  Thomas Jefferson National Accelerator Facility, Newport
News, VA 23606, USA~\footnote{{\href{deurpam@jlab.org}{\tt
deurpam@jlab.org}}}}

\vspace{6pt}

\centerline{Stanley J. Brodsky}

\vspace{3pt}

\centerline {\it SLAC National Accelerator Laboratory, Stanford
University, Stanford, CA 94309,
USA~\footnote{{\href{mailto:sjbth@slac.stanford.edu}{\tt
sjbth@slac.stanford.edu}}}}

\vspace{6pt}

\centerline{Guy F. de T\'eramond}

\vspace{3pt}

\centerline {\it Universidad de Costa Rica, 11501 San Pedro de Montes de Oca, Costa
Rica~\footnote{{\href{mailto:gdt@asterix.crnet.cr}{\tt
gdt@asterix.crnet.cr}}}}

\vspace{10pt}

\begin{abstract}

The QCD running coupling $\alpha_s(Q^2)$ sets the strength of  the interactions of quarks and 
gluons as a function of the momentum transfer $Q$. The $Q^2$ dependence of the coupling is 
required to describe hadronic interactions at  both large and short distances.  In this article we 
adopt the light-front holographic approach to strongly-coupled QCD, a formalism which incorporates 
confinement, predicts the spectroscopy of hadrons composed of light quarks, and describes the  
low-$Q^2$ analytic behavior of the strong coupling $\alpha_s(Q^2)$. The high-$Q^2$ dependence  
of the coupling $\alpha_s(Q^2)$ is  specified by perturbative QCD and its renormalization group equation.  
The matching of the high and low $Q^2$ regimes  of  $\alpha_s(Q^2)$ then determines the scale 
$Q_0$ which sets the interface between perturbative and nonperturbative hadron dynamics. 
The value of $Q_0$ can be used to set the factorization scale for DGLAP evolution of hadronic 
structure functions and the ERBL evolution of distribution amplitudes.  We discuss the  
scheme-dependence of the value of $Q_0$  and the infrared fixed-point  of  the QCD coupling.  
Our analysis is carried out for the  $\overline{MS}$, $g_1$, $MOM$ and $V$ renormalization 
schemes. Our results show that the discrepancies on the value of $\alpha_s$ at large distance 
seen in the literature can be explained by different choices of renormalization schemes. We also 
provide the formulae to compute $\alpha_s(Q^2)$ over the entire range of 
space-like momentum transfer for  the different renormalization schemes discussed in this article.

\end{abstract}

\tableofcontents

\section{Introduction}

The behavior of  the QCD coupling $\alpha_{s}(Q^2) $ at low momentum transfer $Q$ is a central field of study in hadron physics.  Key questions are the analytic behavior of the coupling  in the infrared  (IR), such as whether it exhibits a nonzero IR fixed point and whether it displays conformal-like behavior at low momentum transfers.  Different theoretical approaches to QCD dynamics, such as lattice gauge theory,   Schwinger-Dyson  equations and  light-front holographic  methods  use different definitions of the QCD coupling and effective charges to study $\alpha_{s}(Q^2)$  in the IR domain~\cite{Deur:tbp2016, Prosperi:2006hx}.

Knowing  the strength of the strong coupling $\alpha_{s}$ in the  nonperturbative domain is necessary for understanding fundamental problems in hadron physics, including the mechanisms for color confinement and the origin of gluonic flux tubes within hadrons.  The magnitude of the coupling at low momentum even has impact on high energy phenomena,  such as the amplitude for heavy-quark pair production near threshold~\cite{Brodsky:1995ds} and the magnitude of the $T$-odd Sivers effects in semi-inclusive polarized deep inelastic scattering~\cite{Brodsky:2002cx}.

There is, however,  no consensus  on the  IR behavior of $\alpha_{s}(Q^2)$. The diversity of possible behaviors can be partly traced back to the different definitions of $\alpha_{s}$ in the  nonperturbative domain.  For example, one can define the QCD coupling as an ``effective charge"  from any perturbatively calculable observable~\cite{Grunberg:1980ja}. The various choices for the coupling typically differ from the standard  perturbative definition, such as $\alpha_{\overline{MS}}$, due to the inclusion of   nonperturbative contributions which eliminate an unphysical Landau pole in the physical domain.   
Indeed, the  inclusion of the nonperturbative contributions leads to a modification of the behavior of the coupling in the IR domain.

Studies which simulate a linear confining potential suggest that $\alpha_{s}(Q^{2})$ diverges as $1/ Q^2$ for  $Q^{2}\rightarrow0$~\cite{Richardson:1978bt, Nesterenko:1999np}.   However, this identification is ambiguous, since the linear confining potential for nonrelativistic  heavy quarks in the usual instant form of dynamics~\cite{Dirac:1949cp}  is equivalent, at large separation distances, to a harmonic oscillator potential in the light-front (LF) form of relativistic dynamics~\cite{Dirac:1949cp, Trawinski:2014msa}.  Furthermore, it should be noted that unlike QED, the QCD potential cannot be uniquely  identified with single gluon exchange. Other  approaches suggest that $\alpha_{s}(Q^{2})$ vanishes as $Q^{2}\rightarrow0$~\cite{Boucaud:2008ji}.

In this paper we shall consider the case where $\alpha_{s}(Q^{2})$ becomes constant at low $Q^{2}$~\cite{Brodsky:2010ur, Cornwall:1981zr, Courtoy:2013qca, Aguilar:2001zy}. This behavior, called the ``freezing'' of the coupling  to  a fixed IR value, is automatic if one defines the coupling from an effective charge, and it is thus appealing from physical considerations~\cite{Brodsky:2010ur, Fischer:2006vf}. On simple terms, confinement implies that long wavelengths of quarks and gluons are cutoff at a typical hadronic size. Consequently, the effects of quantum loops responsible for the logarithmic dependence of $\alpha_s$ vanish and $\alpha_s$ should  freeze to a constant value at hadronic scales~\cite{Brodsky:2007hb, Brodsky:2008be}. There are considerable variations in the literature on what should be the freezing value  of the strong coupling --it typically ranges from 0.6 to  $\pi$~\cite{Deur:tbp2016}.  As noted in Ref. \cite{Brodsky:2010ur},  the choice of  renormalization scheme (RS) can explain an important part of the spread in the freezing values  reported in the literature. As we shall show here, an explicit connection between the large-distance confining dynamics of hadronic physics and the short-distance dynamics of quarks and gluons~\cite{Deur:2014qfa} allows one to quantitatively determine this dependence in any RS.

We shall use here the light-front holographic approach to nonperturbative infrared dynamics~\cite{Brodsky:2014yha}. This innovative approach to color confinement allows us  to determine the behavior of the strong coupling in the IR domain~\cite{Brodsky:2010ur}. Using this framework, one can show that the first-order semiclassical approximation to the light-front QCD Hamiltonian is  formally equivalent to the eigenvalue equations in anti-de Sitter (AdS) space~\cite{deTeramond:2008ht, deTeramond:2013it}. This connection also provides a precise relation between the holographic variable $z$ of AdS$_5$ space and the  light-front variable $\zeta$~\cite{deTeramond:2008ht, Brodsky:2006uqa}. For a two-particle bound state the invariant distance squared between the quark and antiquark in the  light-front wavefunction of a meson is defined as $\zeta^2 = x(1-x) b^2_\perp$, where $x =  k^+/P^+$ is the quark's light-front momentum fraction, and $b_\perp$ is the transverse  separation between the $q$ and $ \bar q$. It is also  conjugate to the invariant mass  
$k^2_\perp/x(1-x) $ of  the $q \bar q$  system.

Light-front holography provides a unification of both light-front kinematics and dynamics: the non-trivial geometry of AdS space encodes the kinematical aspects, and the deformation of the action in AdS$_5$ space -- described in terms of a specific dilaton profile $e^{+\kappa^2 z^2}$, encompasses  confinement dynamics and determines the effective potential $\kappa^4 \zeta^2$ in the light-front Hamiltonian~\cite{deTeramond:2013it}. The eigenvalues of the resulting light-front Hamiltonian predict the Regge spectrum of the hadrons, consistent with experiments, and its eigenfunctions determine the light-front wavefunctions underlying form factors, structure functions and other properties of hadrons. The  value of the mass parameter $\kappa$ can be determined from a single hadronic input, such as the proton mass: $\kappa= m_p/2$.

A further advantage of  the light-front holographic mapping is that one can determine the analytic behavior of the strong coupling in the IR: It has the form $\alpha_s(Q^2) \propto  \exp{\left(-Q^2/ 4 \kappa^2 \right)}$. This prediction  follows from the IR modification of AdS space, \emph{i.e.},  from the same dilaton profile which predicts the Regge spectrum~\cite{Brodsky:2010ur, Brodsky:2014yha}.  As we have shown in Ref. \cite{Brodsky:2010ur}, this form gives a remarkable description of the effective charge $\alpha_{g1}(Q^2)$ determined from measurements of the $g_1$ polarized structure function of the nucleon~\cite{Bjorken:1966jh, Deur:2005cf}.

One can also show that the analytic dependence of the confinement potential is uniquely determined by enforcing conformal symmetry --an exact symmetry of the QCD classical Lagrangian when  quark masses are neglected. This method, originally discussed by de Alfaro,  Fubini  and Furlan (dAFF) in the context of one-dimensional quantum field theory allows one to determine uniquely the confinement potential in bound-state equations  while keeping the action conformally invariant~\cite{deAlfaro:1976je}.  One can extend the conformal quantum mechanics of dAFF to 3+1 physical space-time on the light front~\cite{Brodsky:2013ar}. The resulting confinement potential is the transverse harmonic oscillator $\kappa^4 \zeta^2$ in the light-front Hamiltonian which successfully  describes hadronic spectra and form factors~\cite{Brodsky:2014yha}.  Conversely, LF holography determines the AdS$_5$ dilaton profile  $e^{+\kappa^2 z^2}$ and thus the analytic dependence $\alpha_s(Q^2) \propto  \exp{\left(-Q^2/ 4 \kappa^2 \right)}$ of the strong coupling in the IR.

This view has received recently strong support from superconformal quantum mechanics~\cite{Akulov:1984uh, Fubini:1984hf} and its extension to light-front physics~\cite{deTeramond:2014asa, Dosch:2015nwa}. This new approach to hadron physics captures very well the essential physics of QCD confining dynamics and gives remarkable connections between the baryon and meson spectra. Furthermore, it gives remarkable connections  across the full heavy-light hadron spectra, where heavy quark masses break the conformal invariance, but the underlying supersymmetry  still holds~\cite{Dosch:2015bca}. In this framework, the emergent dynamical supersymmetry is not a consequence of supersymmetric QCD, at the level of fundamental fields,  but relies on the fact that in $SU(3)_C$ a diquark can be in the same color representation as an antiquark, namely a $\bf \bar 3 \sim \bf 3 \times \bf 3$.

We shall  show in this paper how the holographic procedure can be extended to describe the strong coupling in the nonperturbative and perturbative domains for any choice of effective charge and RS. The large momentum-transfer dependence  of the coupling $\alpha_{s}(Q^2)$   is  specified by perturbative QCD (pQCD) and its renormalization group equation.  The  matching of the high and low momentum transfer regimes  of $\alpha_{s}(Q^2)$ determines the scale $Q_0$ which sets  the interface between the perturbative and nonperturbative regimes.   Since the value of $Q_0$ determines the starting point for pQCD, it can be used to set the factorization scale for DGLAP evolution of hadronic structure functions \cite{Gribov:1972ri} and the ERBL evolution of distribution amplitudes \cite{Lepage:1979zb}. We will also discuss the dependence of $Q_0$ on the choice of the  effective charge used to define the running coupling and the RS used to compute its behavior in the perturbative regime.   Our analysis also determines the infrared fixed-point behavior of the QCD coupling  as well as the value of the infrared fixed point,  $\alpha_s(0)$,  for any choice of effective charge and RS.

\section{Holographic mapping and matching procedure}

The QCD coupling $\alpha_{s }$ can be defined as an ``effective charge"~\cite{Grunberg:1980ja} satisfying the standard renormalization group evolution equation.  This is analogous to the definition of the QED coupling from the potential between heavy leptons by Gell Mann and Low~\cite{GellMann:1954fq}.  
As we shall show in this section, the analytic behavior of the running coupling $\alpha_{s }$ in the low-$Q^{2}$ nonperturbative domain~\cite{Brodsky:2010ur} can be uniquely predicted   using the light-front holographic approach to strongly coupled QCD~\cite{Brodsky:2014yha}. It can then be matched~\cite{Deur:2014qfa} to form the running coupling at large $Q^2$ as predicted by perturbative QCD in any RS. 

The low $Q^{2}$-evolution of $\alpha_{s}$ is derived from the long-range confining forces:   the originally constant, \emph{i.e.} the conformal invariant light-front
holographic (LFH)  coupling $\alpha_{s}^{LFH}\equiv g_{LFH}^{2}/4\pi$, is 
redefined to include the effects of QCD's long-range confining force.
As we shall show in detail, the  $Q^{2}$ dependence of the coupling in the IR follows from the specific 
embedding of  light-front dynamics in AdS space~\cite{Brodsky:2010ur}; it is uniquely determined in terms of the dilaton profile 
originating from the specific breaking of conformal invariance consistent with the dAFF mechanism~\cite{deAlfaro:1976je, Brodsky:2013ar}.   Likewise, the coupling at short distances, as described by perturbative QCD,  becomes $Q^{2}$-dependent because short-distance QCD quantum effects are included in its definition.

We start with the dilaton modified AdS$_5$ action:
\begin{equation}
S_{AdS}=  - \frac{1}{4}\intop d^4x \, dz\sqrt{\vert g \vert} \,e^{\varphi(z)}\frac{1}{g_{AdS}^{2}} F^2,
\end{equation}
where $g$ is the AdS metric determinant, $g_{AdS}$ the AdS coupling, and the dilaton profile is given  by $\varphi= \kappa^2 z^2$. 
The scale $\kappa$  controls quark confinement and determines the hadron masses in LF holographic QCD~\cite{Brodsky:2014yha}.  
It also determines the $Q^{2}$ dependence of the strong coupling from  the large-distance confining forces, {\it i.e.} the effect of the modification of the AdS space curvature from nonconformal confinement dynamics~\cite{Brodsky:2010ur}:
\begin{equation}
g_{AdS}^{2}\rightarrow g_{AdS}^{2}e^{ -\kappa^{2}z^{2}}.
\end{equation}
The five-dimensional coupling $g_{AdS}(z)$
is mapped,  modulo a  constant, to the  LFH  coupling $g_{LFH}(\zeta)$ of the confining theory in physical space-time.  The holographic variable $z$ is identified with the physical invariant impact separation variable $\zeta$~\cite{deTeramond:2008ht, Brodsky:2006uqa}:
\begin{equation}
g_{AdS}(z) \to g_{LFH}(\zeta).
\end{equation}
We thus have
\begin{equation}  \label{gLFH}
\alpha_s^{LFH}(\zeta) \equiv \frac{g_{LFH}^2(\zeta)}{4 \pi} \propto  e^{-\kappa^2 \zeta^2} . 
\end{equation}

The physical coupling measured at  space-like 4-momentum squared $Q^2=-q^2$ is the light-front transverse Fourier transform
of the  LFH  coupling $\alpha_s^{LFH}(\zeta)$  (\ref{gLFH}):
\begin{equation} \label{eq:2dimFT}
\alpha_s^{LFH}(Q^2) \sim \int^\infty_0 \! \zeta d\zeta \,  J_0(\zeta Q) \, \alpha_s^{LFH}(\zeta),
\end{equation}
in the $q^+ = 0$ light-front frame where $Q^2 = -q^2 = - \mathbf{q}_\perp^2 > 0$, and $J_0$
is a Bessel function. 
Using this ansatz we then have from  Eq.  (\ref{eq:2dimFT}) 
\begin{equation}
\alpha_{s}^{LFH}\left(Q^{2}\right)=\alpha_{s}^{LFH}\left(0\right)e^{-Q^{2}/4\kappa^{2}} \label{eq:alpha_g1 from AdS/QCD}.
\end{equation}

The effective charge  $\alpha_{g_1} = g^2_1/4 \pi$ is defined  from the  integral appearing in the Bjorken sum rule~\cite{Bjorken:1966jh, Deur:2005cf} 
\begin{equation} \label{BjSR}
\frac{\alpha_{g_{1}}(Q^2)}{\pi} = 1 -  \frac{6}{g_{A}} \int_{0}^{1^-} d x\, g_{1}^{p-n}(x,Q^2),
\end{equation}
where $x= x_{Bj}$ is the Bjorken scaling variable, $g_1^{p-n}$  is the isovector component of the nucleon spin structure function, and $g_A$ is the nucleon axial charge.
The IR fixed-point   of $\alpha_{g_1}$  is  kinematically constrained  to  the value $\alpha_{g_1}(0)/\pi = 1$. 
However, we will ignore this constraint here since one of our goals is  to  
determine the freezing value of $\alpha_s(0)$ for  different RS  from the matching procedure described below. 
An agreement with the  value  $\alpha_{s}(0)/\pi  = 1$ for the $g_{1}$-effective charge will demonstrate the consistency of the procedure.

Eq. (\ref{eq:alpha_g1 from AdS/QCD}) is valid only in the  nonperturbative
regime.  However, it can be continued to the pQCD domain thanks to
an overlap existing between the pQCD and nonperturbative QCD regimes
known as parton-hadron duality~\cite{Bloom:1970xb, Melnitchouk:2005zr}.
The  nonperturbative  coupling, Eq. (\ref{eq:alpha_g1 from AdS/QCD}),
and its $\beta$ function,  $\beta(Q^2) = d \alpha_s(Q^2)/ d \log (Q^2)$,  can be equated to their pQCD counterparts for 
each RS considered here.
Thus, we shall impose  the conditions $\alpha_{s}^{pQCD}(Q_{0}^{2})=\alpha_{s}^{LFH}(Q_{0}^{2})$
and $\beta^{pQCD}(Q_{0}^{2})=\beta^{LFH}(Q_{0}^{2})$, where the transition 
scale  $Q_0^2$ indicates the onset of  the pQCD regime as obtained from the matching procedure.  The solution
of this  system of equations is unique,  providing  a relation between
a  nonperturbative quantity, such as $\kappa$ or $\alpha_{s}(0)$,
and the fundamental QCD scale $\Lambda$ in a given RS.  It also sets the value of  $Q_{0}^{2}$. 
This matching procedure was  used  in Ref. \cite{Deur:2014qfa} to determine the QCD scale in the $\overline{MS}$
scheme, $\Lambda_{\overline{MS}}$. It is found that $\Lambda_{\overline{MS}}=0.341\pm0.032$
GeV,  in remarkable  agreement with the combined world data $\Lambda_{\overline{MS}}^{(3)}=0.340\pm0.008$
GeV~\cite{Agashe:2014kda}  and the latest lattice calculations \cite{Vairo:2015vgb}. In this  article, we will use the known
values of $\Lambda$ in several  RS to obtain  the corresponding values of $\alpha_{s}(0)$ and $Q_{0}$.

\section{Results}

In this section we shall derive the form of $\alpha_{s}^{LFH}(Q^{2})$ for effective charges assuming the value $\kappa = 0.51\pm 0.04$ GeV. The value of the RS-independent
scale $\kappa$  
is obtained by averaging  the predictions of light-front holography for the 
$\rho$-meson mass, $\kappa=M_{\rho}/\sqrt{2}$, and the nucleon mass,   
$\kappa=M_{N}/2$~\cite{Brodsky:2014yha}.  The scale $\kappa$ can also be extracted from other observables, including  
hadron masses~\cite{Brodsky:2014yha}, an extension of the holographic model to describe hadron form factors~\cite{Brodsky:2014yha}, and the  low  $Q^2$ dependence of the Bjorken  integral~\cite{Brodsky:2010ur, Deur:2014qfa}.
For example, the determination of $\kappa$ from the  measurements of the Bjorken integral yields   the value $\kappa$ = 0.513 +/- 0.007 GeV.
The $\pm 0.04$ variation covers the possible values of $\kappa$ and is 
characteristic of the uncertainties associated with the
approximations to strongly coupled QCD using the LFH  approach.
The RS-dependent freezing value $\alpha_{s}(0)$ will be  left  as a free parameter.

We shall use the perturbative coupling $\alpha_{s}^{pQCD}(Q^2)$ calculated up to order
$\beta_{3}$ in the perturbative series of the $\beta$ function
\begin{equation} 
Q^{2}\frac{\partial\alpha_{s}}{\partial Q^{2}} =\beta\left(\alpha_{s}\right)=-\left(\frac{\alpha_{s}}{4\pi}\right)^{2}\sum_{n=0}\left(\frac{\alpha_{s}}{4\pi}\right)^{n}\beta_{n}. 
\end{equation}
We take the number of quark  flavors $n_{f}=3$, and the
values of  the QCD scale $\Lambda$ in each scheme as determined at large
$Q^{2}$;   see Table \ref{tab:Different-freezing-values AdS/QCD}. 
The matching procedure then allows us to establish the IR-behavior of $\alpha_{s}$ in any RS.

An approximate analytical expression valid up to order $\beta_{3}$ can be obtained by iteration~\cite{Chetyrkin:1997sg}:
\begin{multline} 
\alpha_{s}^{pQCD}(Q^2) = \frac{4\pi}{\beta_{0}ln\left(Q^{2}/\Lambda^{2}\right)} \biggl[ 1-\frac{\beta_{1}}{\beta_{0}^{2}}\frac{ln\left[ln(Q^{2}/\Lambda^{2})\right]}{ln(Q^{2}/\Lambda^{2})}  \\
+ \frac{\beta_{1}^{2}}{\beta_{0}^{4}ln^{2}(Q^{2}/\Lambda^{2})}\left(\left(ln\left[ln(Q^{2}/\Lambda^{2})\right]\right)^{2}-ln\left[ln(Q^{2}/\Lambda^{2})\right]-1+\frac{\beta_{2}\beta_{0}}{\beta_{1}^{2}}\right) \\
+ \frac{\beta_{1}^{3}}{\beta_{0}^{6}ln^{3}(Q^{2}/\Lambda^{2})}\biggl(-\left(ln\left[ln(Q^{2}/\Lambda^{2})\right]\right)^{3}+\frac{5}{2}\left(ln\left[ln(Q^{2}/\Lambda^{2})\right]\right)^{2} \\
+2 ln\left[ln(Q^{2}/\Lambda^{2})\right]-\frac{1}{2}-3\frac{\beta_{2}\beta_{0}}{\beta_{1}^{2}}ln\left[ln(Q^{2}/\Lambda^{2})\right]+\frac{\beta_{3}\beta_{0}^{2}}{2\beta_{1}^{3}}\biggr)
\biggr] .
\label{eq:alpha_s}
\end{multline}
This result is valid in the $\overline{MS}$ (minimal-subtraction), $V$ (potential) and $MOM$ (momentum-subtraction) schemes \cite{Agashe:2014kda}. 

The first two coefficients of 
the $\beta$ series 
\begin{equation}
\beta_{0}=11-\frac{2}{3}n_{f},
\end{equation}
 and
\begin{equation}
\beta_{1}=102-\frac{38}{3}n_{f},
\end{equation} 
are scheme independent.  The higher order coefficients for the $\overline{MS}$  renormalization scheme are~\cite{Gross:1973id} 
\begin{equation}
\beta_{2} = \frac{2857}{2}-\frac{5033}{18}n_{f}+\frac{325}{54}n_{f}^{2},
\end{equation}
and 
\begin{multline}
\beta_{3} = \left(\frac{149753}{6}+3564 \, \xi\left(3\right)\right)-\left(\frac{1078361}{162}+
\frac{6508}{27}\xi\left(3\right)\right)n_{f}+\left(\frac{50065}{162}+\frac{6472}{81} \, \xi
\left(3\right)\right)n_{f}^{2}   \\
+\frac{1093}{729}n_{f}^{3},
\end{multline}
with the Ap\'{e}ry constant $\xi\left(3\right)\simeq1.20206$. 
In the $MOM$ scheme and Landau gauge, the coefficients are~ \cite{Chetyrkin:2000dq}
\begin{equation}
\beta_{2}=3040.48-625.387n_{f}+19.3833n_{f}^{2},
\end{equation}
 and 
\begin{equation}
\beta_{3}=100541-24423.3n_{f}+1625.4n_{f}^{2}-27.493n_{f}^{3}.
\end{equation}
Four-loop calculations are also available in the related $minimal MOM$ scheme~\cite{Gracey:2013sca}. In the $V$ scheme  the coefficients  are~\cite{Peter:1997me}
\begin{equation}
\beta_{2}=4224.181-746.0062n_f+20.87191n_{f}^{2},
\end{equation}
 and 
\begin{equation}
\beta_{3}=43175.06-12951.7n_f+706.9658n_f^2-4.87214n_{f}^{3}.
\end{equation}
Finally, in the $g_1$ scheme/effective charge, the perturbative coupling expression is \cite{Deur:2005cf}:
\begin{equation}
\label{eq:msbar to g_1}
\alpha_{g_{1}}(Q^2)=\alpha_{\overline{MS}}+3.58\frac{\alpha_{\overline{MS}}^{2}}{\pi}+20.21\frac{\alpha_{\overline{MS}}^{3}}{\pi^{2}}+175.7\frac{\alpha_{\overline{MS}}^{4}}{\pi^{3}} .
\end{equation}
The $\beta$-series order for the $\alpha_{\overline{MS}}$ expression in Eq.~(\ref{eq:msbar to g_1}) is typically taken to be the same as the $\alpha_{\overline{MS}}$ order of Eq.~(\ref{eq:msbar to g_1});  this is 4$^{\mbox{\footnotesize{th}}}$ order in the present case.

\begin{figure}[ht]
\centerline{\includegraphics[width=.5\textwidth]{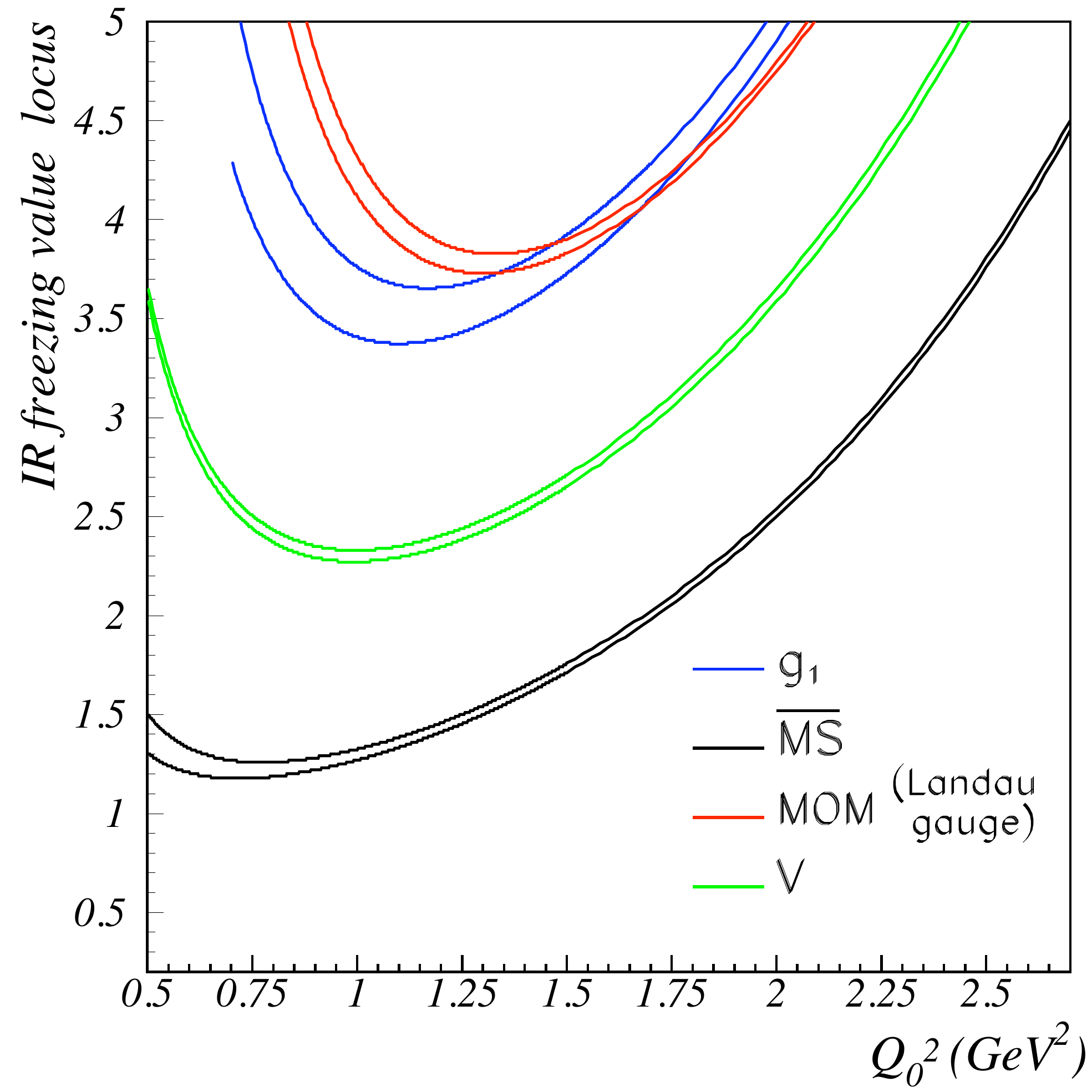}}
\caption{\label{Q0} \small The freezing value $\alpha_{s}(0)$ versus
the transition scale $Q_{0}^{2}$. Calculations are done in $\overline{MS}$
scheme (black lines), the $g_{1}$ scheme (blue lines), the $V$ scheme
(green lines) and the MOM scheme (red lines). Two lines of same color
represent results obtained with $\alpha_{s}^{pQCD}$ calculated either
at order $\beta_{2}$ or $\beta_{3}$. The extrema of these curves
provide the value of $Q_{0}^{2}$ and $\alpha_{s}(0)$ that meets the
matching conditions $\alpha_{s}^{pQCD}(Q_{0}^{2})=\alpha_{s}^{LFH}(Q_{0}^{2})$
and $\beta^{pQCD}(Q_{0}^{2})=\beta^{LFH}(Q_{0}^{2})$. }
\end{figure}

\begin{figure}[ht]
\centerline{\includegraphics[width=.5\textwidth]{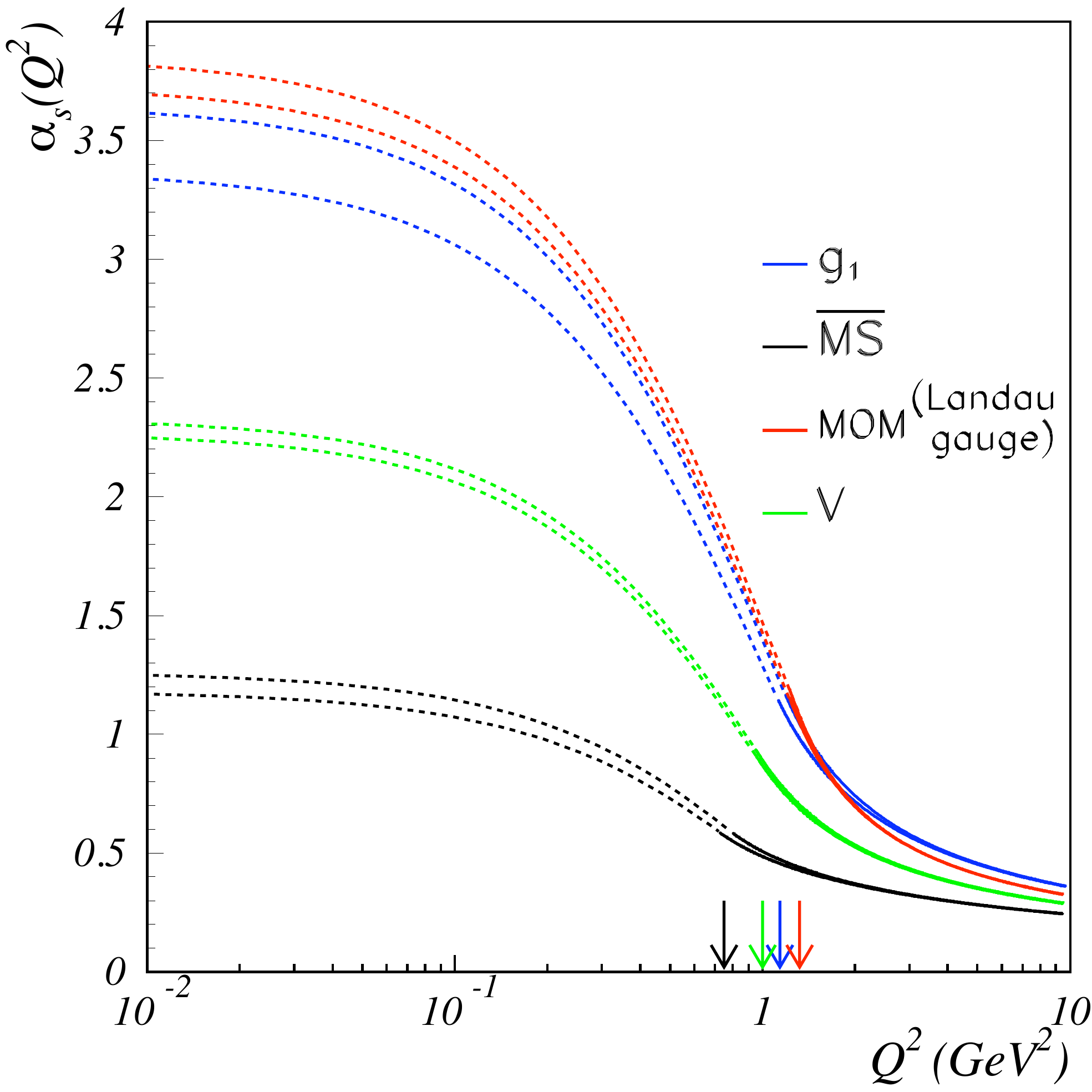}}
\caption{\label{Flo:different freezings} \small The strong coupling $\alpha_{s}(Q^{2})$
for different schemes. The continuous lines are the perturbative calculations
done either at order $\beta_{2}$ or $\beta_{3}$. The dashed curves
are their matched continuations into the non-perturbative domain.
The location of the  scale  $Q_0^2$ for the transition from the nonperturbative to the perturbative region is shown  by the arrows for each scheme.}
\end{figure}

\begin{figure}[ht]
\centerline{\includegraphics[width=.5\textwidth]{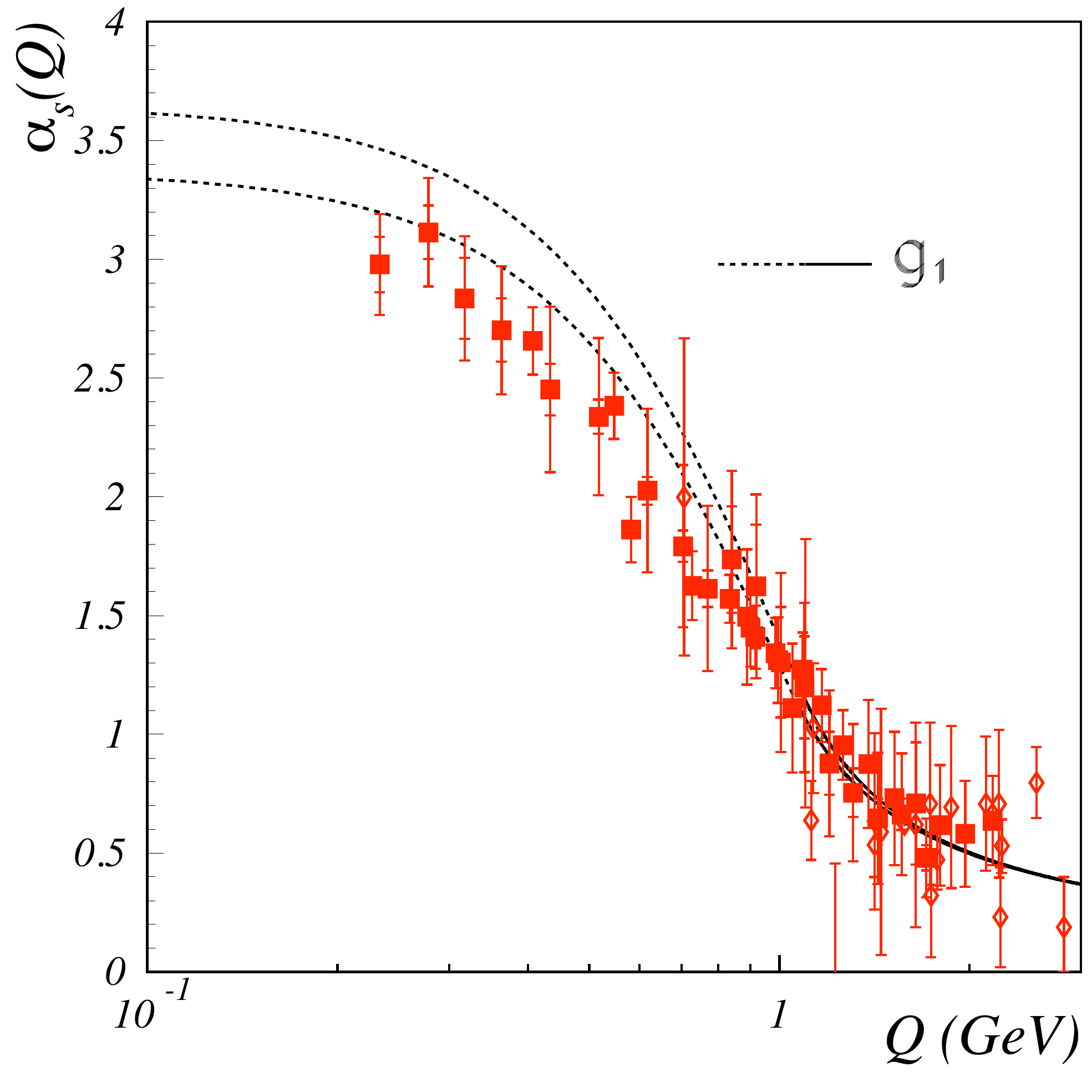}}
\caption{\label{Flo:comp} \small
Comparison  between the experimental data~\cite{Deur:2005cf} and the prediction from our matching procedure and the value of $\Lambda$ listed in Table \ref{tab:Different-freezing-values AdS/QCD}. The coupling is calculated in the $g_1$ scheme. The inner error bar on each experimental data point is the point-to-point uncorrelated uncertainty and the outer error bar represents the total uncertainty (point-to-point correlated and uncorrelated uncertainties added in quadrature). }
\end{figure}

We have carried out  the matching procedure numerically for the $\overline{MS}$, $V$, $MOM$ (choosing the Landau gauge) and the $g_{1}$ schemes. 
Our results are presented in Figs.  \ref{Q0}, \ref{Flo:different freezings} and \ref{Flo:comp}. 
In Fig. \ref{Q0}, we show $\alpha_{s}(0)$
as a  function of $Q_{0}^{2}$ for the first matching condition $\alpha_{s}^{pQCD}(Q_{0}^{2})=\alpha_{s}^{LFH}(Q_{0}^{2})$.
The two curves illustrated  for each  scheme represent the results when the
matching is done with $\alpha_{s}^{pQCD}$ calculated either at order
$\beta_{2}$ or $\beta_{3}$. For the $g_{1}$ scheme, the expression
of $\alpha_{g_{1}}^{pQCD}$ is a series in $\alpha_{\overline{MS}}^{pQCD}$
rather than in $\beta_{i}$, see Eq. (\ref{eq:msbar to g_1}). In that case the calculations are done
at fourth order in $\alpha_{\overline{MS}}^{pQCD}$ calculated at
$\beta_{2}$ or $\beta_{3}$. 

The second matching condition requires
the continuity of the $\beta$-function, $\beta^{pQCD}(Q_{0}^{2})=\beta^{LFH}(Q_{0}^{2})$.
The solution is given by the extrema of the curves. 
The two matching conditions provide the
values of $\alpha_{s}(0)$ and $Q_{0}^{2}$.  The corresponding couplings
are shown in Fig. \ref{Flo:different freezings}. 
A comparison between data and our result
for $\alpha_{g_1}(Q^2)$ is shown on Fig. \ref{Flo:comp}.

The difference between
the results obtained with $\alpha_{s}^{pQCD}$ calculated at order
$\beta_{2}$ or  at order $\beta_{3}$ provide an estimate of the uncertainty
due to the series truncation.   Other contributions (not shown in the
figures) come from the uncertainties in the values of $\kappa$ and  $\Lambda$.
For  the latter, we have assumed a 5\% relative uncertainty. 

The different
freezing values and $Q_{0}^{2}$ scales obtained are listed in Table
\ref{tab:Different-freezing-values AdS/QCD}.
Our results for $\alpha_{s}(0)$  can be compared to the typical values from the literature. 
Most of the results from Lattice QCD, the Schwinger-Dyson formalism, stochastic quantization and the
functional renormalization group equations are carried out in the $MOM$ scheme and 
Landau gauge for $n_{f}=0$ ~\cite{Deur:tbp2016}. These computations
yield $\alpha_{s}(0)=2.97$~\cite{Deur:tbp2016},
which can be  compared 
with $\alpha_{s}(0)=2.84$ obtained using  our procedure with $n_{f}=0$. 
The result from Cornwall~\cite{Cornwall:1989gv} using $n_f=3$ and in the $\overline{MS}$ scheme
yields $\alpha_{s}(0)=0.91$~\cite{Deur:tbp2016}, in agreement  with our $\overline{MS}$ determination.
The constraint  $\alpha_{s}(0)=\pi$ in the $g_1$ scheme also agrees well with our analysis.

\begin{table}
\begin{center}
\begin{tabular}{|c|c|c|c|}
\hline 
$\alpha_{s}(0)$ & RS & $Q_{0}^{2}$ (GeV) & $\Lambda$ (GeV)\tabularnewline
\hline
\hline 
$1.22\pm0.04\pm0.11\pm0.09$ & $\overline{MS}$ & $0.75\pm0.03\pm0.05\pm0.04$ & $0.34\pm0.02$\tabularnewline
\hline 
$2.30\pm0.03\pm0.28\pm0.21$ & $V$ & $1.00\pm0.00\pm0.07\pm0.06$ & $0.37\pm0.02$\tabularnewline
\hline 
$3.79\pm0.06\pm0.65\pm0.46$ & $MOM$ & $1.32\pm0.02\pm0.10\pm0.08$ & $0.52\pm0.03$\tabularnewline
\hline 
$3.51\pm0.14\pm0.49\pm0.35$ & $g_{1}$ & $1.14\pm0.04\pm0.08\pm0.06$ & $0.92\pm0.05$\tabularnewline
\hline
\end{tabular}
\caption{\small \label{tab:Different-freezing-values AdS/QCD}Freezing values for
$\alpha_{s}$ (column 1) calculated in different  schemes (column 2) and for $n_f=3$.
The scale  of the pQCD onset is given in the third column. 
The  RS-dependent  values of $\Lambda$ are in the fourth column.
The first, second and third uncertainties on $\alpha_{s}(0)$ and
$Q_{0}^{2}$ stem from the truncation of the $\beta$-series determining
$\alpha_{s}^{pQCD}$, the $\pm0.04$  GeV uncertainty on $\kappa$ and
the 5\% uncertainty on $\Lambda$, respectively. }
\end{center}
\end{table}

The  scheme dependence of the freezing value is easily understood by
considering  the slope of $\alpha_{s}$ near $Q_{0}^{2}$, which depends
on the scheme-dependent value of $\Lambda$: the steeper the slope,
the larger $\alpha_{s}(0)$. The  scheme dependence of the transition
scale $Q_{0}^{2}$ is likewise easily explained: the smaller the
freezing value, the earlier the onset of pQCD. Our $Q_{0}^{2}$ values
are close to the value found in Ref. \cite{Courtoy:2013qca}  
where, in order to explain parton-hadron duality, the evolution of
$\alpha_{\overline{MS}}^{pQCD}(Q^{2})$ near $Q_{0}^{2}\simeq1$ GeV$^{2}$
is stopped. It is also consistent with the transition value $Q_{0}^{2}=0.87$
GeV$^{2}$ found in Ref.~\cite{Gomez:2015tqa} using the
$\overline{MS}$  scheme.

The  full $Q^2$ dependence of $\alpha_s(Q^2)$ for a specific RS  can be conveniently represented in the form
\begin{equation}
\alpha_s(Q^2) = \alpha_s(0)e^{\frac{-Q^2}{4 \kappa^2}}H(Q^2-Q_0^2) + \bigl(1-H(Q^2-Q_0^2)\bigr)\alpha_s^{pQCD}(Q^2),
\end{equation}
where $\kappa=0.51\pm0.04$ GeV, $H(Q^2)$ is the Heaviside step function, $\alpha_s^{pQCD}$ is 
given by Eq.~\ref{eq:alpha_s} for the $\overline{MS}$, $MOM$ or $V$ schemes or by Eq.~(\ref{eq:msbar to g_1}) for the $g_1$ scheme, and $\alpha_s(0)$ and $Q^2_0$ are given in Table~\ref{tab:Different-freezing-values AdS/QCD}.

\section{Conclusions}

The dependence of the freezing
value of $\alpha_{s}(Q^{2})$ at low $Q^{2}$  on the choice of the effective charge and the pQCD renormalization scheme can be quantitatively
estimated using the  light-front holographic approach to strongly coupled QCD and the matching
procedure described in Ref. \cite{Deur:2014qfa}.  The  results we have obtained in this paper for $\alpha_{s}(Q^{2}=0)$ in the deep infrared,  ranging from 0.98 to 4.96,
show that 
the choice of renormalization scheme and the choice of the effective charge used to define the QCD coupling strongly influences its freezing
value. 
For example, the freezing values reported in the literature typically range
from   $\sim 0.6$ to $\sim 3$: accounting for the scheme/effective charge dependence  thus
resolves a large part of this discrepancy.  In fact, our values of $\alpha_{s}(0)$
for the $\overline{MS}$, $MOM$ and $g_1$ schemes agree with the corresponding typical values 
encountered in the literature.

Other factors must also be considered before comparing 
various couplings proposed in the literature, including which approximations are used.  For example,
many calculations in the $MOM$ scheme are done without dynamical
quarks.  If one  takes $n_{f}=0$ we find a central value $\alpha_{s}(0)=2.84$ in the $MOM$ scheme, 
in comparison with $\alpha_{s}(0)=3.79$ for $n_{f}=3$. 
Another factor is the choice of gauge for gauge-dependent definitions of $\alpha_{s}(Q^{2})$, such as the ones defined from vertices and propagators~\cite{Deur:tbp2016} or the definition using the gluon self-energy and the pinch technique~\cite{Cornwall:1981zr, Binosi:2009qm}. It was however shown that an appropriately chosen gauge can lead to similar behavior between the pinched defined coupling and the one defined from the ghost-gluon vertex~\cite{Aguilar:2009nf}. This was demonstrated in the Landau gauge and the $MOM$ scheme. Different couplings can be defined from other vertices and propagators, but they are related to the ghost-gluon vertex coupling. These relations have been discussed in Ref.~\cite{Gracey:2011vw} for the $MOM$ scheme.

As we have shown, matching the high and low momentum transfer regimes  of the running QCD coupling, as determined from light-front holographic QCD and pQCD evolution, determines the scale $Q_0$ which sets the interface between perturbative and nonperturbative hadron dynamics.   Above $Q_0$, the perturbative gluon and quark degrees of freedom are relevant. Below $Q_0$, the collective effects of the gluonic interactions can be understood to provide the potential $\kappa^4 \zeta^2$ in the effective LF Hamiltonian underlying light quark meson and baryon spectroscopy.   In addition,  the collective gluonic effects can provide the basis for phenomena such as the  ``flux tubes"~\cite{Isgur:1984bm,Bjorken:2013boa}, which are postulated to connect the incident quark and diquarks in high energy hadronic collisions.

The specific numerical value for the transition scale $Q_0$ is also important for hadron physics phenomenology.   The value of $Q_0$ can be understood as the starting point for pQCD  evolution from gluonic radiation;  it thus can be used to set the factorization scale for DGLAP evolution of hadronic structure functions \cite{Gribov:1972ri} and the ERBL evolution of distribution amplitudes \cite{Lepage:1979zb}.

The use of the transition scale $Q_0$ to eliminate the factorization scale uncertainty,  in combination with the ``principle of maximum conformality" (PMC)~\cite{Wu:2013ei},  which sets renormalization scales order by order  to obtain scheme-independent pQCD predictions for observables, can eliminate important theoretical uncertainties and thus  greatly improve the precision of pQCD predictions for collider phenomenology.

\section*{Acknowledgments}

This material is based upon work supported by the U.S. Department of Energy, Office of Science, Office of Nuclear Physics under contract DE--AC05--06OR23177. This work is also supported by the Department of Energy  contract DE--AC02--76SF00515. SLAC-PUB-16449.

\end{document}